\documentclass[fleqn,usenatbib,useAMS]{mnras}

\usepackage{graphicx}	
\usepackage{amsmath}	
\usepackage{amssymb}	

\title[CP stars in NGC\,2516]{
The evolutionary state of the chemically peculiar members of the open cluster NGC\,2516
 }

\stepcounter{footnote}

\author[Kharchenko et al.]{
N.~V.~Kharchenko$^{1}$\thanks{Corresponding author: nvkhar@gmail.com},
A.~E.~Piskunov$^{2}$,
S.~Hubrig$^{3}$,
M.~Sch\"oller$^{4}$
\\
$^{1}${Main Astronomical Observatory, National Academy of Sciences of Ukraine, 27 Akademika Zabolotnoho St, 03143 Kyiv, Ukraine} \\
$^{2}${Institute of Astronomy, Russian Academy of Sciences, 48 Pyatnitskya St, Moscow 119017, Russia} \\
$^{3}${Leibniz-Institut f\"ur Astrophysik Potsdam (AIP), An der Sternwarte~16, 14482~Potsdam, Germany} \\
$^{4}${European Southern Observatory, Karl-Schwarzschild-Str.~2, 85748 Garching, Germany} \\
}

\date{Accepted XXX. Received YYY; in original form ZZZ}

\pubyear{2022}

\begin{document}
\label{firstpage}
\pagerange{\pageref{firstpage}--\pageref{lastpage}}
\maketitle

\begin{abstract}
We aim at establishing safe membership and evolutionary status of 11 chemically peculiar 
(CP) stars that are residing in the 
domain of the open cluster NGC~2516 and are frequently referred to as cluster members.
We queried the Gaia EDR3 catalogue in an
area with a radius of 1\,deg and selected 37508 stars brighter than $G=19$\,mag. The cluster membership was 
determined in parallax-proper motion-space and 719 probable and 764 possible members were found. 
The obtained average astrometric and photometric parameters of the cluster
are in good agreement with the most recent literature data.
The evolutionary status of the target 
stars was determined with respect to Padova isochrones. 
After minor adjustments
including the metallicity, the reddening, and the transformation scale variation,
a perfect fit of the model to the observations 
over the whole observed magnitude range was achieved.
Only 5 of the 11 considered CP stars could be classified as highly probable 
cluster members.
Among the Ap/Bp stars with previously detected magnetic fields
HD\,65987 and HD\,65712 have a high membership probability and the magnetic star CPD$-$60\,944B is a 
possible cluster member.
Further we discuss the blue straggler nature of HD\,66194 and the magnetic star HD\,65987.
To our knowledge, HD\,65987 is currently the only known blue straggler,
with a field of the order of a few hundred Gauss.
The most striking result of our study is that the strongly magnetic A0p star 
HD\,66318 with previously reported very low fractional age does not belong to the NGC\,2516 cluster at a high level
of confidence. 
\end{abstract}




\section{Introduction}
\label{sec:intro}

Knowledge about the evolutionary state of the intermediate-mass main-sequence chemically peculiar (CP)  A- and B-type 
stars is essential to understand both the physical processes taking place in these stars and the origin of 
their magnetic fields. \citet{moss03} suggested that magnetic Ap/Bp stars acquire their magnetic field at 
the time of their formation or early in their evolution and what is currently observed is then a fossil 
field. The competing dynamo theory proposes that the magnetic field is generated by a turbulent dynamo 
operating in the star's convective core \citep{2002A&A...381..923S}.
As long as it was accepted that strong magnetic fields are 
observed at all evolutionary states from the zero-age main sequence (ZAMS) to the terminal-age main 
sequence (TAMS), one of the difficulties for the dynamo theory was to explain how the field reaches the 
stellar surface in the rather short time available before the arrival of the star on the main sequence. 
Alternatively, magnetic fields may be generated by strong binary interaction, i.e., in stellar mergers. 
Scenarios for the origin of the magnetic fields of Ap/Bp stars, in which these stars result from the merging 
of two lower mass stars or protostars were suggested by \citet{2010ARep...54..156T} and 
\citet{2009MNRAS.400L..71F}, respectively. The mergers would produce a brief period of strong differential 
rotation and give rise to large-scale fields in the radiative envelopes. It is therefore possible 
that the binaries with Ap/Bp components that we observe now were triple systems earlier in their history
(e.g. \citealt{2017A&A...601A..14M}). 

Obviously, understanding the origin and evolution of the magnetic fields of Ap/Bp stars requires knowledge 
of their evolutionary status. Whether they become magnetic at a certain evolutionary state before reaching 
the ZAMS, or during core hydrogen burning, or at the end of their main-sequence life requires either systematic 
studies of field stars with accurate Gaia parallaxes, established cluster or association members with known 
ages, or binary systems.

In previous studies of the evolutionary state of magnetic Ap and Bp stars with accurate Hipparcos parallaxes 
\citep{1997A&A...323L..49P} and photometric data in the Str\"omgren or the Geneva system 
\citep{1993A&A...269..403H,1996BaltA...5..303H}, \citet{2000ApJ...539..352H,2005ASPC..343..374H} showed 
that the distribution of magnetic stars with masses below 3\,$M_\odot$ differs from that of normal stars in 
the same temperature range at a high level of significance. Normal A stars occupy the whole width of the main 
sequence, without a gap, whereas magnetic stars are concentrated towards the centre of the main-sequence band. 
On the other hand, \citeauthor{2007A&A...470..685L} (\citeyear{2007A&A...470..685L}, and references therein) 
studied about 80 Ap/Bp stars that are potential members of open clusters with masses in the range between 2 
and 10\,$M_\odot$. In contrast to the results of \citet{2000ApJ...539..352H,2005ASPC..343..374H}, the authors 
reported that magnetic fields are present at essentially all evolutionary stages between ZAMS and TAMS for 
stellar masses between about 2 and 5\,$M_\odot$. 

More recent studies based on very different datasets imply that a certain amount of time is necessary for 
the magnetic field to build up to become measurable using spectropolarimetry. These studies, such as the 
volume-limited survey of 52 Ap and Bp stars within 100\,pc by \citet{2019MNRAS.483.3127S}, the survey of 
294 Ap/Bp stars by \citet{2020MNRAS.493.3293B}  using a colour-magnitude diagram based on the homogeneous 
Gaia DR2 photometry from \citet{2018A&A...616A..17A}, and the LAMOST DR4 survey by \citet{2018A&A...619A..98H}, 
confirm the concentration of Ap/Bp stars with $M<3\,M_\odot$ towards the centre of the main-sequence band. 

A  substantial number of potential CP stars were previously identified in several open clusters
using photometric and spectroscopic surveys.
Compared to other open clusters, the open cluster NGC\,2516 is well-known to harbor a comparatively large 
number of chemically peculiar stars (e.g.\ https://webda.physics.muni.cz), including classical Ap/Bp stars 
with measured magnetic fields and stars with a HgMn peculiarity, possessing only rather weak magnetic fields 
\citep[e.g.][]{1995A&A...293..810M,1995ComAp..18..167H,2012A&A...547A..90H,2020MNRAS.495L..97H}. Furthermore, 
the cluster NGC\,2516 is especially interesting in view of the frequent detection of X-ray sources among 
chemically peculiar stars. As reported by \citet{1997MNRAS.287..350J}, chemically peculiar late-B and A stars 
are more likely to be detected as X-ray emitters than normal A-type stars at a confidence level of 90--95\%, 
although it is not clear whether this emission is intrinsic and magnetospheric in origin, or if the observed 
X-ray emission is generated in unresolved late-type companions. As only very few magnetic CP stars are known 
to be members of close binaries, it is quite possible that the observed emission is intrinsic. Assuming an 
intrinsic model for the generation of the X-ray emission, magnetically confined wind-shocks were considered by 
\citet{1997A&A...323..121B}. 
Among the identified chemically peculiar late-B and A stars in NGC\,2516, X-ray 
emission was detected in HD\,66318, CPD$-$60\,978, CPD$-$60\,981, HD\,66295, the binary CPD$-$60\,944, and 
HD\,65949. Only the magnetic stars HD\,65712 and HD\,65987 remained undetected in the survey 
of \citet{1997MNRAS.287..350J}. 

Measurable magnetic fields have been reported for five Ap and Bp stars assumed to be cluster members: 
HD\,66318, HD\,65987, CPD$-$60\,944B, HD\,66295, and HD\,65712  \citep{2006A&A...450..777B}. Among them, 
HD\,66318 (A0p SrCrEu), the coolest of these five stars, possesses a strong mean longitudinal magnetic 
field with a strength of about 4.5\,kG, and has been 
studied in detail by \citet{2003A&A...403..645B}.  Using the isochrone for a cluster of age of 
1.6$\times10^{8}$\,yr and the Geneva stellar evolution tracks for $Z=0.02$ from \citet{1992A&AS...96..269S}, 
the authors concluded that HD\,66318 has completed only about $16\pm5$\% of its main 
sequence life. This conclusion contradicted the results by \citet{2000ApJ...539..352H} that magnetic fields 
appear in Ap stars after about 30\% of their main sequence lifetime has elapsed. 

Advantageously, the recent 
availability of homogeneous Gaia EDR3 data allows us now to reconsider diverse characteristics of the open 
cluster NGC\,2516 and to put much more reliable constraints on membership and age of the chemically peculiar stars.
The first successes of space astrometry presented by the Hipparcos project invoked a series of proper motion 
and photometric membership all-sky studies of open clusters and led to the identification of a uniform population 
of the Galactic disk \citep{2005A&A...438.1163K,2013A&A...558A..53K}. The recent highly accurate and homogeneous 
Gaia EDR3 release \citep{2021A&A...649A...1G} caused a further rise of publications on studies of galactic 
open clusters and on stars populating these clusters.  \citet{2021arXiv211101819C} reported the identification of 664 new 
clusters used as main tracers of the Milky Way 
spiral structure. Gaia EDR3 data were used by \citet{2021arXiv211105291T} to study the external regions 
(coronae and tidal 
tails) of 467 local open clusters. The authors found extended coronae in 389 objects and identified footprints of 
tidal tails in 71 of them. Similarly, \citet{2021arXiv211004296H} studied in the Gaia EDR3 data the wide neighbourhood of 
the nearest young clusters (Pleiades, $\alpha$~Per, NGC~2451A, IC~2391, IC~2502) reporting 1700 confident 
members in the inner parts of the clusters and 1200 candidate members outside. \citet{2021RNAAS...5..173L} 
and \citet{2021ApJ...912..162P} used Gaia EDR3 to study the 3D morphology of open clusters and found 
evidence of tidal tails in 
nearby clusters. \citet{2022MNRAS.509.1664J} joined Gaia EDR3 astrometric and Gaia-ESO spectroscopic data for 
the calculation of the 3D kinematic membership probability for 63 open and 7 globular clusters.
There are also recent 
publications devoted exclusively to the NGC\,2516 cluster. A study of the wide neighbourhood of this cluster using
 Gaia DR2/EDR3, TESS, Gaia-ESO and GALAH data aiming at the confirmation 
of the existence of a 500\,pc wide halo was carried out by \citet{2021AJ....162..197B}. 
\citet{2021ApJ...923...23H} used spectrophotometric observations 
combined with data from the TESS, Gaia-ESO, and GALAH surveys to study stellar rotation in NGC~2516.

\begin{figure*}
   \centering
\includegraphics[width=0.90\hsize,clip=]{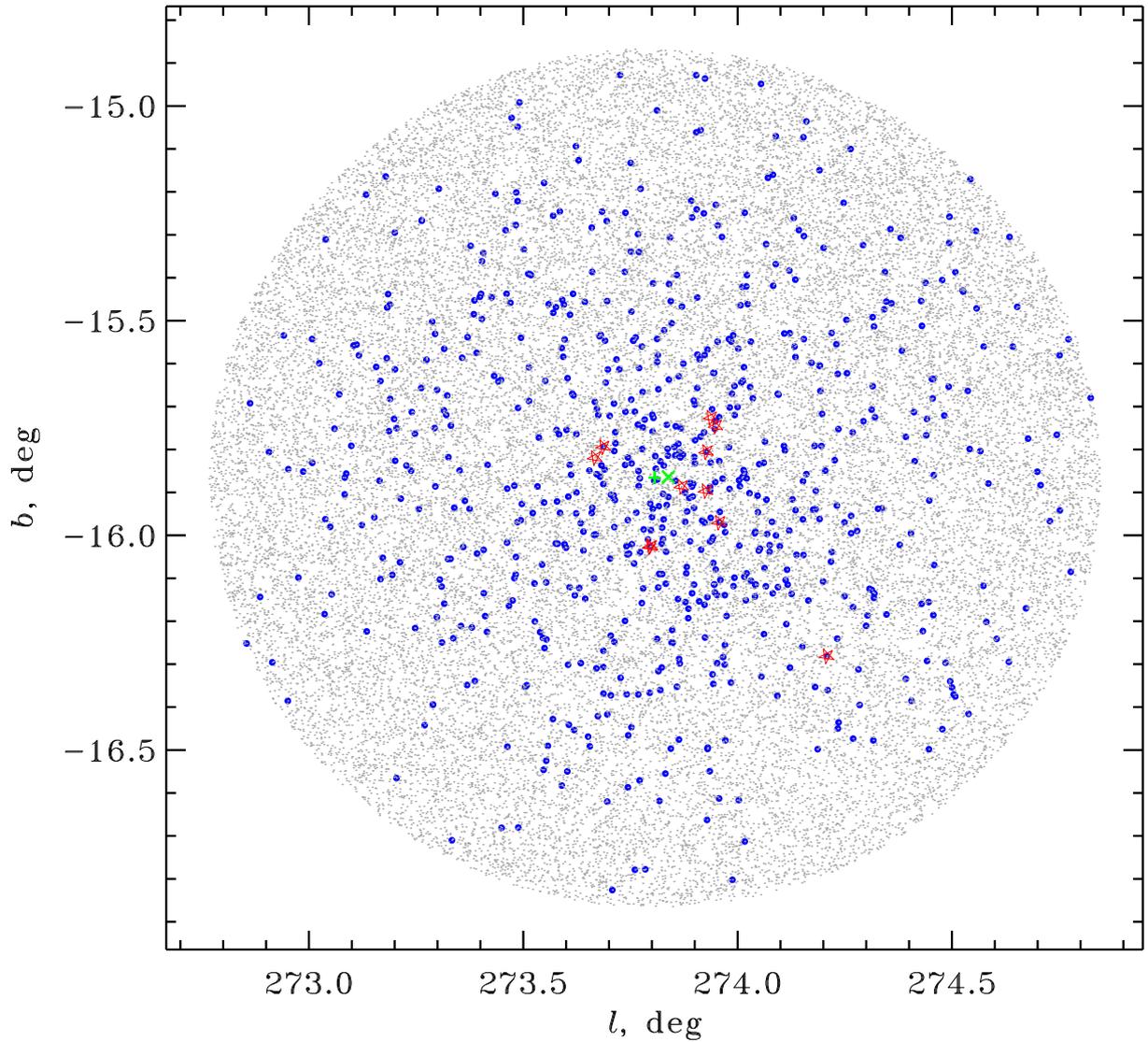}
\caption{Geometry of the sky area with the stars queried from Gaia EDR3. Grey dots mark all queried objects and
blue dots correspond to the selected cluster members as described in Sect.~\ref{sec:memb}. The green symbols indicate the 
adopted cluster centre (the plus corresponding to the Milky
Way Star Clusters \citep{2013A&A...558A..53K} and the cross corresponding to the current average).
Red five-point stars 
represent the CP targets selected for this study (see Sect.~\ref{sec:data}). }
\label{fig:skyarea}
\end{figure*}

In this paper, we 
concentrate on the determination of Gaia EDR3-based cluster membership and respective evolutionary status of 
the known CP stars observed in the area of NGC~2516.
In the following sections, we show our analysis of the astrometric and astrophysical parameters of NGC~2516 and discuss 
the membership probability of the currently known CP stars as well as of a few presumably normal targets. 
Finally, we discuss the impact of our results 
on the current understanding of the magnetic field origin in such stars.

\section{Characterising NGC\,2516 with Gaia EDR3}\label{sec:cluparam}

The primary goal of this study is to examine on the basis of Gaia EDR3 data \citep{2021A&A...649A...1G} 
the cluster membership of known CP stars in NGC~2516.
Since the CP stars in our sample are bright and occupy the cluster central area, there is no need 
to consider the data completeness all over the cluster area, nor the full representativity of the 
cluster member sample. On the other hand, in order to derive membership correctly, based on average parameters of 
the cluster members, a wider coverage of the cluster population is necessary. Therefore, the general approach of this 
study is to obtain secure cluster membership on the basis of the most accurate average parameters of the cluster by 
taking into account the astrometric accuracy limits, the photometric limits due to faint stars and  
the spatial distribution avoiding the cluster outskirts.

\subsection{Input sample}\label{sec:data}

The data were queried with the \textit{topcat} facility from the Gaia EDR3 catalogue\footnote{tables 
\texttt{gaiaedr3.gaia\_source} and \texttt{gaiaedr3.gaia\_source\_cor\-rec\-ti\-ons}} at the ARI 
site\footnote{https://gaia.ari.uni-heidelberg.de/tap}. The query area is centered at the cluster 
position ($\textit{RA}=119.490$, $\textit{Dec}=-60.750)$\,deg as defined in the Milky
Way Star Clusters (MWSC) project \citep{2013A&A...558A..53K}, 
with a radius equal to 1\,deg, which according to the MWSC exceeds the apparent radius of the cluster by a factor of 
two. We applied corrections for the parallax, the $G$ magnitude, and the flux excess
\citep{2021A&A...649A...3R,2021A&A...649A...4L} using the algorithms 
from \url{https://www.cosmos.esa.int/web/gaia/edr3-code} and the data provided by ARI's Gaia services. 
The corrections for saturation for the few brightest sources ($G<8$\,mag) are applied according to 
\citet{2021A&A...649A...3R}. In total 88815 objects were selected, 8635 of them were rejected due to
lack of data in at least one of the 26 queried columns. Considering the astrometric and photometric limitations
mentioned above, the final working sample consisted of 37508 objects with a brightness level $G<19$\,mag. 
The search area and the queried stars are shown in Fig.~\ref{fig:skyarea}

The sample of CP stars previously reported as cluster members 
includes five Ap/Bp stars with measurable magnetic fields, 
HD\,66318, HD\,65987, CPD$-$60\,944B, HD\,66295, and HD\,65712, the two HgMn stars HD\,65949 and HD\,65950, and 
the blue straggler candidate fast rotating B3Vn star HD\,66194 (e.g.\ \citealt{2000AJ....119.2296G}).
Three more targets were included in our sample: CPD$-$60\,944A, CPD$-$60\,978, and CPD $-$60\,981.
The visual companion in the system CPD$-$60\,944, CPD$-$60\,944A, exhibits in the spectra strong
lines of \ion{Hg}{ii}, \ion{Mn}{ii}, \ion{P}{ii}, \ion{Ga}{ii}, and \ion{Xe}{ii}
and was identified as a HgMn star by \citet{2014MNRAS.443.1523G}.
The chemical peculiarities in the atmosphere of CPD$-$60\,978, typical for magnetic Ap stars, were mentioned 
for the first time by \citet{1972A&A....21..373D}. The target CPD $-$60\,981 is
a short-period 
eclipsing binary with $P_{orb}=3.2$\,d \citep{2001A&A...374..204D} and was classified as an Ap star with 
SrCrEu peculiarity type by \citet{1976ApJ...205..807H}. 
All three targets were reported as X-ray emitters, although no definite detection of a magnetic field was reported for 
them by \citet{2006A&A...450..777B}.

\begin{figure}
   \centering
\includegraphics[width=0.95\hsize,clip=]{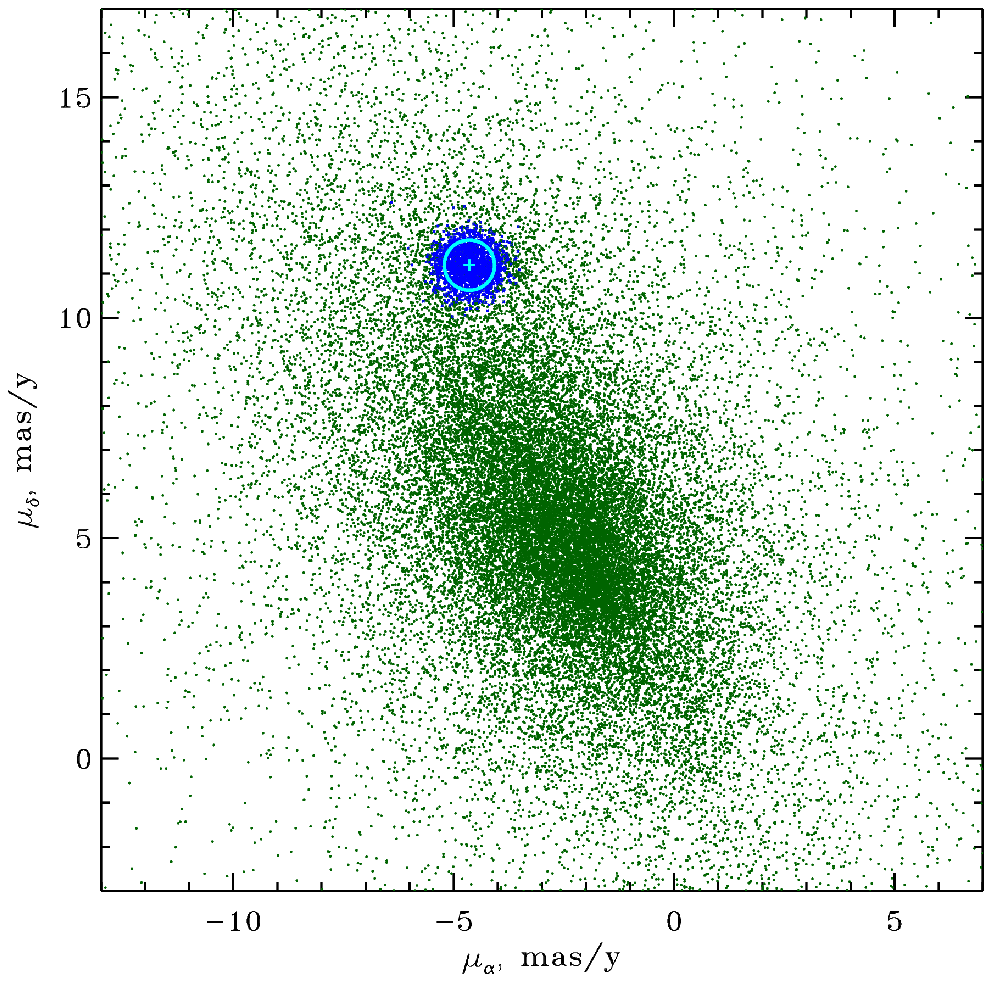}
\includegraphics[width=0.95\hsize,clip=]{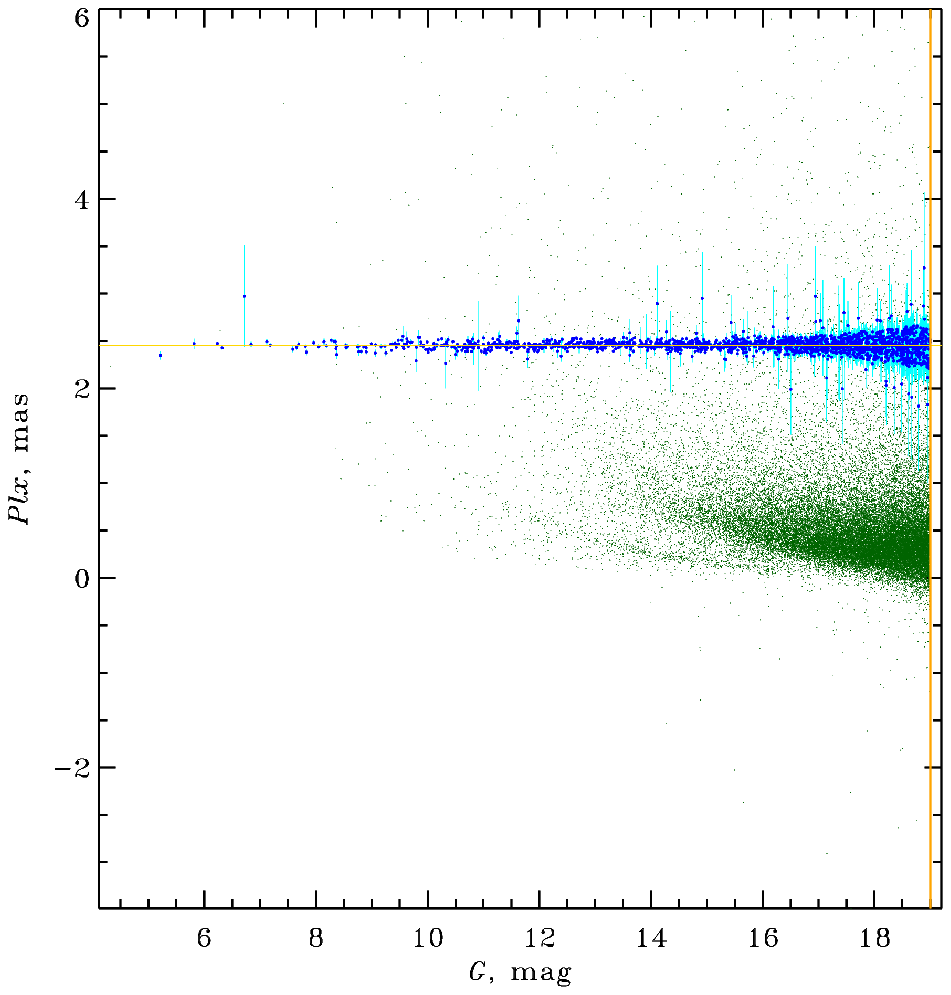}
\caption{
The distribution of stars in the NGC~2516 area in the vector point (top panel) and parallax-magnitude 
(bottom panel) diagrams. Green dots show field stars, blue dots indicate stars used for the calculation of 
the cluster member distribution parameters, where the cyan plus corresponds to the average proper motion. The 
orange horizontal line indicates the average parallax. The large cyan oval (almost a circle) indicates the 
ellipse of the proper motion standard deviations $\sigma_{\mu_{\alpha}}$ and $\sigma_{\mu_{\delta}}$. The cyan 
bars in the bottom panel show individual parallax errors for cluster members. The vertical orange line is 
the adopted magnitude limit.
}
\label{fig:vpdplx}
\end{figure}

\subsection{Cluster membership}\label{sec:memb}

Since the primary goal of this work is to study the evolutionary status of the Ap cluster member stars, which is
directly related to the cluster color-magnitude-diagram (CMD),
we decided, to avoid unnecessary biases, not to use 
photometric member selection, which normally is a part of the  MWSC membership pipeline \citep{2012A&A...543A.156K}. 
In addition, as EDR3 has not yet released measurements of radial velocities (RVs), 
in order to keep the kinematic membership approach uniform over all studied objects, we do 
not use in our study the RV data for membership evaluation. This is one of the basic differences between
our work and the works of \citet{2007A&A...470..685L} and \citet{2000AJ....119.2296G}, who also considered radial 
velocities. Given the proximity of the cluster, it is sufficient to select cluster members based  
exclusively on proper motion and parallax, which allows us to identify among 
the selected members a number of co-moving field stars.

Following \citet{2012A&A...543A.156K}, our study makes use of a probabilistic selection of cluster 
members. Each star in the dataset is assigned the value $P_{c}$, representing a combined probability of sharing space and 
kinematics with other members:
\begin{equation}
\label{eq:ptot}
    P_{c}=\min\{P_{\rm kin},\,P_{\rm \varpi}\},
\end{equation}
where the probabilities based on kinematics (essentially proper motions) and parallax data, 
$P_{\rm kin}$ and $P_{\rm \varpi}$, are calculated as described below.
Since the search area covers a relatively restricted area of the sky, the kinematic probability $P_{\rm kin}^{i}$ for the $i$-th 
star to belong to the cluster is defined in its simplified form
\begin{equation}\label{eq:pkin}
P_{\rm kin}^{i}=
 \exp\left\{-\left[\left(\frac{\mu_{\alpha}^{i}-\overline{\mu}_{\alpha}}{2\,\sigma_{\mu_{\alpha}}}\right)^2+
 \left(\frac{\mu_{\delta}^{i}-\overline{\mu}_{\delta}}{2\,\sigma_{\mu_{\delta}}}\right)^2  \right] \right\}, \\    
\end{equation}
where $\mu_{\alpha}^{i}$ and $\mu_{\delta}^{i}$ are values of proper motions of the star in right 
ascension\footnote{here and further we denote $\mu_{\alpha}=\mathrm{d}\alpha/\mathrm{d}t\,\cos \delta$} 
and declination, and $\overline{\mu}_{\alpha}$ and $\overline{\mu}_{\delta}$ are average proper motions, 
corresponding to the centre of the member distribution in the vector point diagram, the VPD 
(see Fig.~\ref{fig:vpdplx} for illustration). The parameters 
$\sigma_{\mu_{\alpha}}$ and $\sigma_{\mu_{\delta}}$ are the observed standard deviations, arising both from internal 
motions of the cluster stars and from the measurement errors. In the case of NGC~2516, the former effect is of 
the order of 0.1--0.5\,mas/yr, which is much higher than the observation errors $\varepsilon_{\mu}$, which in the 
magnitudes range of interest ($G<17$\,mag) 
are much lower ($\varepsilon_{\mu}<0.05$ mas/yr), and are 
comparable to it at fainter magnitudes only.
The distribution parameters were calculated in an iterative process in a close neighbourhood
of the distribution center,
rapidly converging to the final values. The initial guess parameters were selected by eye.

A visual inspection of the diagram $(\varpi,G)$ (see e.g. Fig.\ref{eq:pkin}) indicates a perfect separation of 
cluster member candidates and field stars. Therefore we decided to apply also for the calculation of
$P_{\varpi}$ a simple approach:  
\begin{equation}
P_{\varpi}^{i}=\exp\left\{-\left(\frac{\varpi^{i}-\overline{\varpi}}{2\,\sigma_{\varpi}}\right)^2\right\}, 
\end{equation}
assuming that the distribution parameters are the same as in Eq.~(\ref{eq:pkin}). Here $\varpi^{i}$ is 
the measured parallax of $i$th star, $\overline{\varpi}$ is the member average parallax, and $\sigma_{\varpi}$ 
is the standard deviation of the distribution. In contrast to the proper motions, 
the dispersion of the parallaxes is dominated by the inaccuracy of the observations rather than by the actual 
dispersion of the distances within the cluster. 

Following MWSC \citep{2012A&A...543A.156K}, the above probability distribution parameters are used for 
cluster membership classification. We adopt the criterion that stars shifted from the distribution centre by less 
than one standard deviation ($P\gtrsim0.61$) are highly probable proper motion or parallax members of the 
cluster, also referred to as $1\sigma$-members.
We consider the $1\sigma$-stars as bona fide cluster members.
Those stars with deviations between one and two standard deviations 
($0.14\lesssim P\lesssim0.61$) are called possible (or $2\sigma$) members, and compose together with $1\sigma$-stars 
the bulk of the potential member population. The next category of outliers (3$\sigma$-stars, 
$0.01\lesssim P\lesssim 0.14$) includes a very low fraction of cluster stars, which are classified as possible 
field stars.

\begin{figure}
   \centering
\includegraphics[width=0.95\hsize,clip=]{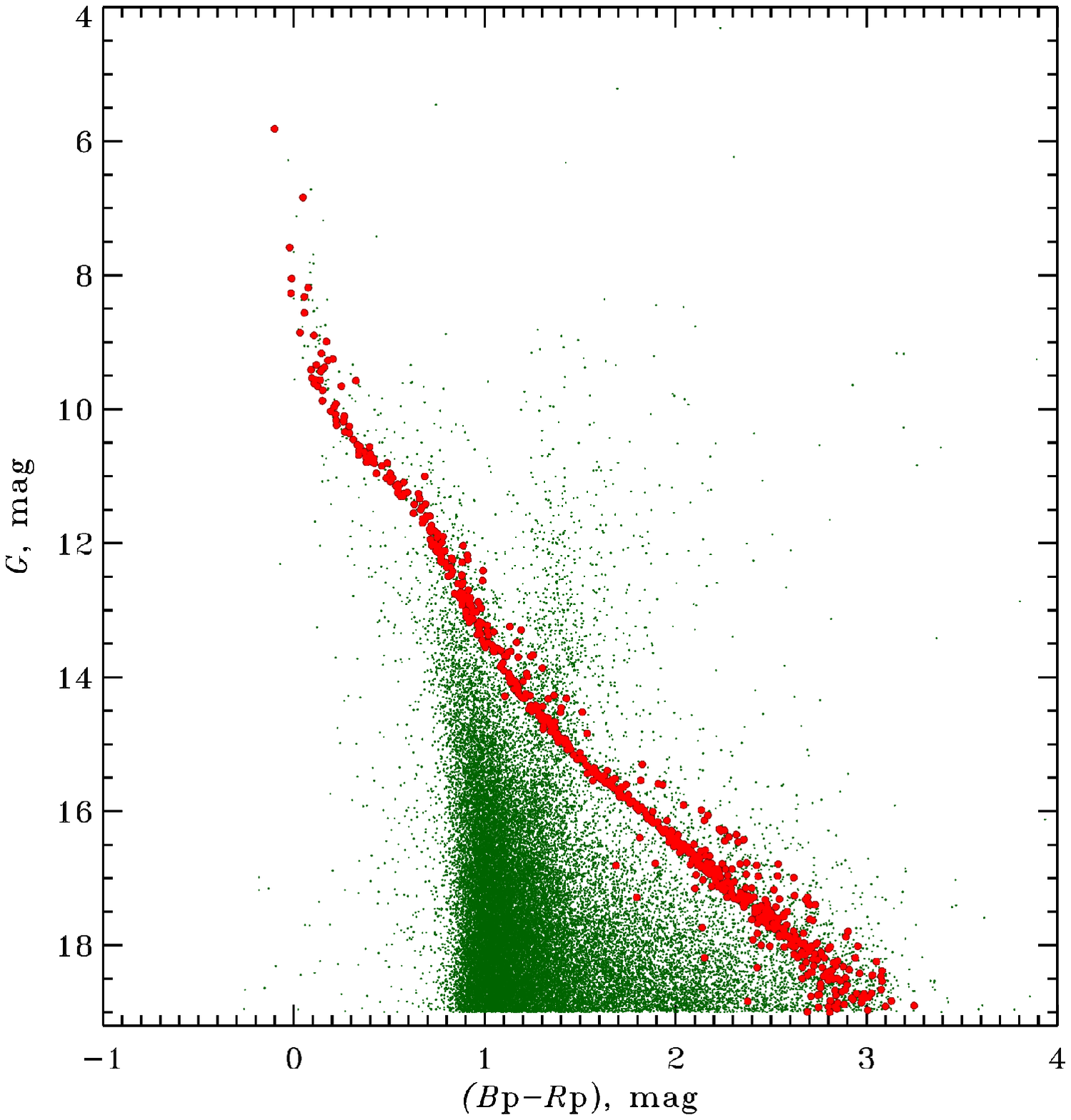}
\includegraphics[width=0.95\hsize,clip=]{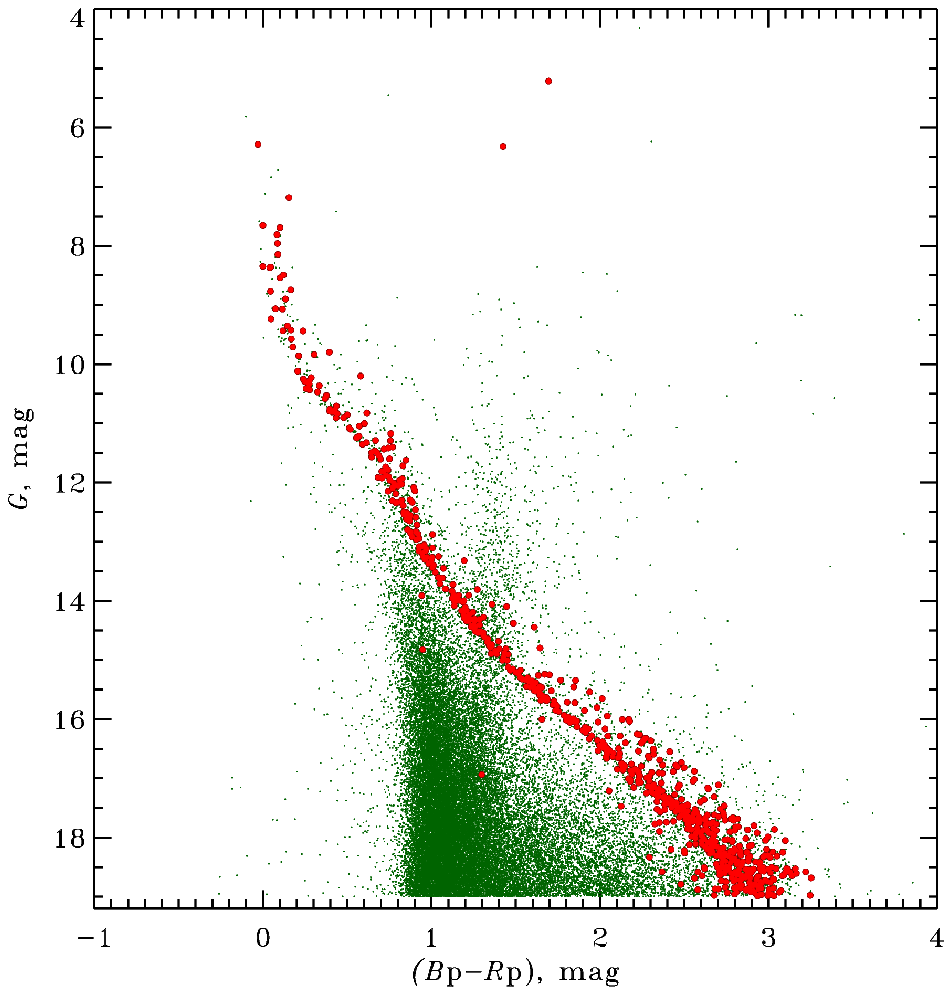}
\caption{
Comparison of the colour-magnitude diagrams of two membership classes.
The top panel is for $1\sigma$ (most probable) cluster members,
and  the bottom panel is for $2\sigma$ members.
Green dots are field stars whereas red dots are member stars.
}
\label{fig:mmb}
\end{figure}

In Fig.~\ref{fig:mmb} we compare the CMDs for two  member classes, the $1\sigma$ and $2\sigma$ members. One can 
see that, in general, both 
populations produce very similar patterns, especially for brighter stars ($G\lesssim 15$\,mag). At fainter 
magnitudes, both sequences demonstrate an increase of the contamination by field stars. As
expected, the contamination degree for the $2\sigma$-population is higher. As a result, we can conclude that 
for the bright magnitude domain the contamination in both diagrams is similar and not too strong. 
At $G>16$\,mag both sequences become broader due to the increase of photometric errors. 

\subsection{Cluster parameters}\label{sec:prm}

As we discuss in Sect.~\ref{sec:memb} (see also Fig.~\ref{fig:mmb}), it is possible to construct an extremely accurate 
CMD for the cluster members, from bright stars down to faint stars, allowing us to investigate the detailed 
structure of the cluster CMD.

Isochrones for the Gaia EDR3 passbands are interpolated with help of \textit{CMD3.5}, the publicly available 
Padova webserver\footnote{\url{http://stev.oapd.inaf.it/cmd}}. An important reference line corresponding to 
the Zero Age Main Sequence (ZAMS) was 
computed as the blue edge of a set of isochrones of various ages and other common parameters (metallicity, 
BC-scale, etc.) extending from brightest to faintest absolute magnitude of interest. 
Since we are interested in the
detailed structure of the CMD, we tried to select the best fitting the isochrone by modest 
variations of isochrone parameters provided by the server around the expected values. 

\begin{figure}
   \centering
\includegraphics[width=0.95\hsize,clip=]{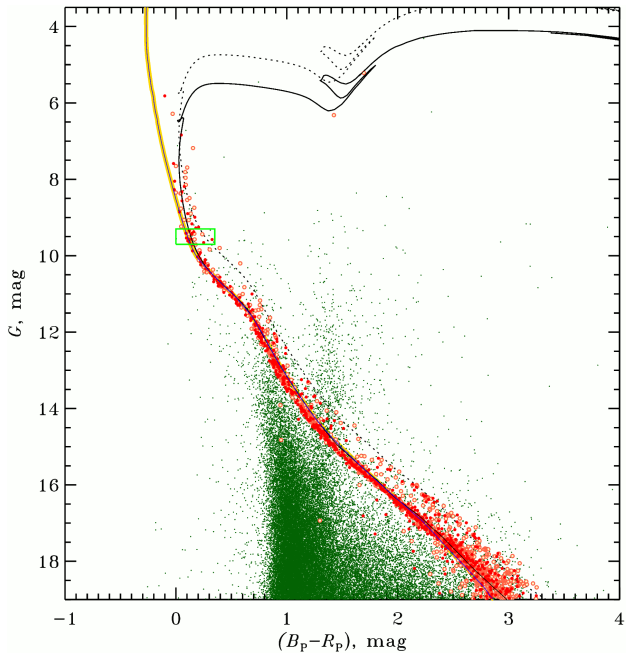}
    \caption{
Observed colour-magnitude diagram of NGC~2516
and the Padova isochrone for $\log t=8.10$ (black line).
The black-yellow line indicates the ZAMS, the dotted line is a sequence of equal-mass unresolved binaries.
All relevant parameters are taken from Table~\ref{tab:cluprm},
using [M/H] = +0.05 and the bolometric corrections scale \textit{YBC}.
The green dots are field stars, the red dots are cluster $1\sigma$-members,
the red open circles mark $2\sigma$-members.
The green rectangle shows an insert enlarged in Fig.~\ref{fig:cmdto}. 
}
\label{fig:cmdmmb}
\end{figure}

Taking into account the spectroscopic evidence of a slightly enhanced metal abundance of the
NGC~2516 stars with respect to the Sun \citep{2014A&A...561A..93H}, we varied the isochrone metallicity between [M/H]\,$=-0.05$ 
and $+0.05$. For the bolometric correction and effective temperature scales, we considered all three options
available in \textit{CMD3.5}. For the determination of reddening, we followed the \textit{CMD3.5} recommendations 
to use the extinction coefficients $A_\lambda/A_0$ for the Gaia photometric bands \citep{2018A&A...616A..10G} 
computed at every isochrone point (except for the OBC case, where one uses constant coefficients). For the
calculation of the scaling extinction $A_0$, we used a relation between the average colour excess 
$E(B_{\rm{P}}-R_{\rm{P}})$ and the extinction $A_{G}$  following the relations by \citet{1989ApJ...345..245C} 
and \citet{1994ApJ...422..158O}, which led to $A_{G}/E(B_{\rm{P}}-R_{\rm{P}}) \approx 2.05$. As a 
reference, we employed the main sequence inflection point at $(G,B_{\rm{P}}-R_{\rm{P}})\approx (11.4,0.65)$. 
The transformation 
of intrinsic isochrone photometry to reddened values is performed along with the following relations:
\begin{equation}
(B_{\rm{P}}-R_{\rm{P}}) = (B_{\rm{P}}-R_{\rm{P}})_0 + E(B_{\rm{P}}-R_{\rm{P}})\,,  \label{eq:redd}
\end{equation}
and
\begin{equation}
G = M_{G} + 10 - 5\,\log\overline{\varpi} + A_G\,,  \label{eq:extn}
\end{equation}
where $(B_{\rm{P}}-R_{\rm{P}})_0$ and $M_{G}$ are the intrinsic colour and the absolute magnitude of the isochrone, 
and respective reddening or extinction are depending on the adopted reddening scheme.

\begin{table}
 \caption{Derived cluster parameters of NGC 2516.}
 \label{tab:cluprm}
\centering
\begin{tabular}{lrlr}
\hline
$\overline{\mu}_{\alpha}$ (mas/yr)  & $-4.648$ & $N_{\rm mmb}$                    & 1483 \\
$\overline{\mu}_{\delta}$ (mas/yr)  & $11.208$ & $E(B_{\rm{P}}-R_{\rm{P}})$ (mag) & 0.18 \\
$\sigma_{\mu_{\alpha}}$ (mas/yr)    & $0.408$  & $A_{G}$ (mag)                    & 0.369\\
$\sigma_{\mu_{\delta}}$ (mas/yr)    & $0.397$  & $G-M_{G}$ (mag)                  & 8.420\\
$\overline{\varpi}$ (mas)           & $2.454$  & $\log t$ (yr)                    & 8.10 \\
$\sigma_{\varpi}$ (mas)             & $0.066$  & $d$ (pc)                         & 407.5\\
\hline
\end{tabular}
\end{table}

As expected, the metal abundance variations within the considered limits only symbolically affect 
the cluster CMD. Therefore, we accept for the isochrone the spectroscopic determination of $\rm{[M/H]}=+0.05$ 
without anticipating any noticeable change of the resulting cluster parameters. The effect of the bolometric 
scales is more prominent. Generally, the comparison of the Gaia observations and the Padova isochrones shows a reasonable 
agreement of the sequences. At the same time, at a lower significance level one observes in some 
temperature/luminosity segments a disagreement between the isochrone and the observations, whereas in 
other segments of the same isochrone the agreement is perfect. 
For example, the isochrones, denoted in the
\textit{CMD3.5} interface as \textit{OBC}, use scales traditional for the Padova server
\citep{2002A&A...391..195G,2017ApJ...835...77M}, 
and perfectly fit the cluster sequence at brighter 
magnitudes ($G \lesssim15$\,mag).
In contrast, the \textit{YBC} and \textit{YBC/Vega} scales are perfect in the bright and the faint 
segments ($G<12$\,mag or $G>16$\,mag),  but disagree with observations at intermediate 
magnitudes. 
Since in this study we are interested in the brighter stars, 
we select the \textit{YBC} scale for further 
comparisons, keeping in mind this peculiarity. 

\begin{table*}
 \caption{Parameters of NGC~2516 from recent open cluster surveys. In the first six columns we present the name of 
the survey, the number of  NGC~2516 members, the average proper motions and parallaxes with their errors 
(columns 3,5,7), standard deviations (columns 4,6,8), the colour excess, the apparent distance modulus,
the distance, log age, and the photometric system. 
}
\label{tab:litprm}
\setlength{\tabcolsep}{2pt}
\centering
\begin{tabular}{rrccccccccccl}
\hline
 \multicolumn{1}{c}{Survey}                  & 
 \multicolumn{1}{c}{N}                       &
 \multicolumn{1}{c}{$\mu_{\alpha}$}          & 
 \multicolumn{1}{c}{$\sigma_{\mu_{\alpha}}$} &  
 \multicolumn{1}{c}{$\mu_{\delta}$}          &  
 \multicolumn{1}{c}{$\sigma_{\mu_{\delta}}$} &  
 \multicolumn{1}{c}{$\varpi$}                &           
 \multicolumn{1}{c}{$\sigma_{\varpi}$}       &
 \multicolumn{1}{c}{$CE$}                    & 
 \multicolumn{1}{c}{$(m-M)$}                 &
 \multicolumn{1}{c}{$d$}                     &  
 \multicolumn{1}{c}{$\log t$}                &
  Photom.\\
 \multicolumn{1}{r}{}         & 
 \multicolumn{1}{r}{}         &
 \multicolumn{1}{c}{(mas/yr)} & 
 \multicolumn{1}{c}{(mas/yr)} &  
 \multicolumn{1}{c}{(mas/yr)} &  
 \multicolumn{1}{c}{(mas/yr)} &  
 \multicolumn{1}{c}{(mas)}    &           
 \multicolumn{1}{c}{(mas)}    &
 \multicolumn{1}{c}{(mag)}    & 
 \multicolumn{1}{c}{(mag)}    &
 \multicolumn{1}{c}{(pc)}     &  
 \multicolumn{1}{c}{(yr)}     &
 \\
  \hline
COCD$^a$&$53$ &$-4.13\pm0.41$  &$-$     &$10.15\pm0.22$  &$-$     & $-$           &$-$    &$0.07$ &$7.91$       &$346$  &$8.08\pm0.12$ & BV  \\
HRR$^b$ &$11$ &$-4.17\pm0.11$  &$-$     &$11.91\pm0.11$  &$-$     &$2.92\pm0.10$  &$-$    &$-$    &$7.68\pm0.07$&$-$    &$-$           & Hp   \\
MWSC$^c$&$698$&$-3.50\pm0.29$  &$-$     &$10.20\pm0.29$  &$-$     &$-$            &$-$    &$0.034$&$7.88$       &$373$  &$8.48\pm0.12$ & JHKs\\
GDR2$^d$&$798$&$-4.748\pm0.017$&$0.441$ &$11.221\pm0.014$&$0.345$ &$2.417\pm0.002$&$0.045$&$-$    &$-$          &$408.9$&$-$           & Gaia \\    
\hline
\multicolumn{12}{l}{$^a$\citet{2005A&A...438.1163K}; $^b$\citet{2009A&A...497..209V}; $^c$\citet{2013A&A...558A..53K}; $^d$\citet{2018A&A...618A..93C}}
\end{tabular}
\end{table*}

In Fig.~\ref{fig:cmdmmb}, we show the observed CMD together with the fitted isochrone and respective 
ZAMS, adopting a metallicity $+0.05$ and using the BC-scale \textit{YBC}. The isochrone is shifted in the CMD 
according to Eqs.~\ref{eq:redd} and \ref{eq:extn} with the extinction coefficients depending on the effective 
temperature and the scaling factor equal to the average extinction that is adopted to be a free parameter. The best 
fit is provided by the average colour excess value $E(B_{\rm{P}}-R_{\rm{P}})=0.18$\,mag and the age $\log t=8.10$. 
One can see that the isochrone perfectly fits the observations in the bright star domain $G<12$\,mag  and for 
faint stars with $G>16$\,mag. For intermediate magnitudes, the isochrone is redder by about 0.1\,mag at maximum. 
We relate this peculiarity to the effect of the BC-scale discussed above: the \textit{OBC} scale provides in 
the intermediate region  of magnitudes a perfect fit. It is worth to note that the selected isochrone 
accurately fits the data both for very bright ($G<7$\,mag) and very faint magnitudes ($G>17$\,mag) where 
one can easily distinguish the Pre-Main-Sequence branch of the cluster, which forks of the ZAMS at $G\approx 17$\,mag. 
The Main Sequence (MS) between $G\approx 9$\,mag and $G\approx17$\,mag is very sharp and well populated, therefore 
conclusions on the isochrone shape along a wide range of magnitudes seem to be very firm. Even for the brightest 
stars in the MS turn off region and in the red supergiant region the agreement is impressive,
in spite of the very poor member statistics.
We also note that apart from the classical population building up the MS, 
there is quite a fraction of stars populating the MS strip with a width of about 0.75\,mag all over 
the magnitude range. Evidently, this fraction of stars is related to unresolved binary/multiple systems, 
either physical or apparent. 
The most impressive example of such stars is the brightest cluster member blue straggler candidate 
HD~66194, evidently deviating from the isochrone, but spatially and kinematically appearing as a 98\% cluster member.

In Table~\ref{tab:cluprm} we present the cluster parameters determined from the cluster member sample constructed 
in Sects.~\ref{sec:memb} and \ref{sec:prm}.

\subsection{Comparison with previous results}\label{sec:compprm}

As a number of studies of NGC\,2516 were published in the post-Hipparcos era, we can compare our 
cluster parameters with those presented in the literature.
There was a series of all-sky star cluster surveys based on Hipparcos observations and collecting 
astrometric and general-purpose information based on homogeneous data from  the
Hipparcos\footnote{https://cdsarc.cds.unistra.fr/viz-bin/cat/I/239} and 
Tycho\footnote{https://cdsarc.cds.unistra.fr/viz-bin/cat/I/259} catalogues. For example, the Tycho-based 
proper motions and photometric membership  established by \citet{2005A&A...438.1163K} provided average 
proper motion, photometric distance and reddening for about 650 local open clusters within about 2 kpc 
from the Sun (the survey is referred hereafter as COCD). \citet{2009A&A...497..209V} has re-reduced the original 
Hipparcos observations\footnote{https://cdsarc.cds.unistra.fr/viz-bin/cat/I/311} and was able to present 
more accurate data (including trigonometric parallaxes) for the 20 nearest open clusters.
The open cluster survey MWSC \citep{2013A&A...558A..53K} was based on ground-based observations 
(the catalogues PPMXL\footnote{https://cdsarc.cds.unistra.fr/viz-bin/cat/I/317}, and 
2MASS\footnote{https://cdsarc.cds.unistra.fr/viz-bin/cat/II/246}) and only indirectly was related to 
the Hipparcos project being astrometrically tied to the Hipparcos/Tycho system. The photometry data of MWSC 
involved 2MASS near-IR photometry \citep{2010AJ....139.2440R}. The most recent open cluster survey 
of \citet{2018A&A...618A..93C} was based on the Gaia DR2 catalogue and demonstrated a revolutionary raise of the 
data quality, compared to previous observations. 

All surveys mentioned above included NGC~2516, one of the clusters nearest to the Sun, allowing us to compare 
their results, presented in Table~\ref{tab:litprm}. The comparison indicates 
a considerable improvement of the cluster parameter data in the study based on the Gaia results. 
At the same time, studies based on different Gaia releases reproduce approximately similar results on the cluster 
position and kinematics. 
A few recent papers discussed the cluster parameters of NGC~2516 in the studies of different 
issues of the nearby open clusters. For example, \citet{2021A&A...645A..84M} used Gaia EDR3 considering a wide 
area five-parameter based cluster membership ($N=1860$)
and reported that the distance to the maximum of the 
member density distribution for NGC~2516 is equal to $d_c=413.8$\,pc. \citet{2021ApJ...912..162P} studied 
the 3D geometry and kinematics of 13 nearby ($d<500$ pc) clusters. The authors applied a wide area cluster member 
identification based on the Gaia EDR3 five-parameter sample ($N=2690$)  and found that the distance to the 
maximum of the member density distribution is equal to $d_c=410.5\pm 3.1$ pc. Using Gaia EDR3 
astrometry and Gaia-ESO spectroscopy, \citet{2022MNRAS.509.1664J} were able to determine the 3D kinematics 
in 70 open and globular clusters and to assign a kinematic membership probability for the studied stars. 
Using highly probable members, they established parallax-based distance moduli and reddenings, which for 
NGC~2516 are equal to $m_0-M=8.07$ and $E(B-V)=0.11$. Employing high quality 
Gaia astrometry and GALAH/APOGEE spectroscopy, \citet{2021MNRAS.503.3279S} determined the metallicity for 134 clusters with 
secure membership. For NGC~2516, from the study of two secure cluster  members, the authors reported a cluster metallicity equal 
to $\mathrm{[Fe/H]}=-0.079\pm 0.025$. In a somewhat earlier work, \citet{2018AJ....156..165C} used UBV data, a few sets of 
isochrones for [Fe/H]\,$=0$, Gaia DR2 proper motions and parallax membership to determine
the age $\log t=8.22-8.29$ and a photometric distance of NGC~2516 $V_0-M_V=8.04-8.11$.

\section{NGC~2516 membership and evolutionary state of the CP stars}\label{sec:indiv}

\begin{table*}
 \caption{
List of the stars investigated for membership together with the Gaia EDR3 parameters and their errors. In the first column we show 
the sample running number, then the EDR3 identification and coordinates, followed by value/error pairs for the $G$-magnitude, 
colour $(B_{\rm{P}}-R_{\rm{P}})$, proper motions $\mu_{\alpha}$, $\mu_{\delta}$, and parallax $\varpi$.
}
 \label{tab:cpstars}
\setlength{\tabcolsep}{2pt}
\centering
\begin{tabular}{rlrrrrrrrrrrrr}
\hline

\multicolumn{1}{c}{\#}                                     &
\multicolumn{1}{c}{Gaia EDR3}                              &
\multicolumn{1}{c}{$l$}                                    &
\multicolumn{1}{c}{$b$}                                    &
\multicolumn{1}{c}{$G$}                                    &
\multicolumn{1}{c}{$\varepsilon_{G}$}                      &
\multicolumn{1}{c}{$B_{\rm{P}}-R_{\rm{P}}$}                &
\multicolumn{1}{c}{$\varepsilon_{(B_{\rm{P}}-R_{\rm{P}})}$}&
\multicolumn{1}{c}{$\mu_{\alpha}$}                         &
\multicolumn{1}{c}{$\varepsilon_{\mu_{\alpha}}$}           &
\multicolumn{1}{c}{$\mu_{\delta}$}                         &
\multicolumn{1}{c}{$\varepsilon_{\mu_{\delta}}$}           &
\multicolumn{1}{c}{$\varpi$}                               &
\multicolumn{1}{c}{$\varepsilon_{\varpi}$}                 \\
\multicolumn{1}{c}{ }                                      &
\multicolumn{1}{c}{ }                                      &
\multicolumn{1}{c}{(deg)}                                  &
\multicolumn{1}{c}{(deg)}                                  &
\multicolumn{1}{c}{(mag) }                                 &
\multicolumn{1}{c}{(mag)}                                  &
\multicolumn{1}{c}{(mag)}                                  &
\multicolumn{1}{c}{(mag)}                                  &
\multicolumn{1}{c}{(mas/y)}                                &
\multicolumn{1}{c}{(mas/y)}                                &
\multicolumn{1}{c}{(mas/y)}                                &
\multicolumn{1}{c}{(mas/y)}                                &
\multicolumn{1}{c}{(mas)}                                  &
\multicolumn{1}{c}{(mas) }                                 \\
\hline
 1 & 5290820682661822848 & 273.687004   & $-$15.793607 & 7.5863 &0.0008 & $-$0.0220 & 0.0011&  $-$5.023 & 0.037&  11.201 & 0.031& 2.414 & 0.031\\
 2 & 5290767390708069888 & 273.937409   & $-$15.725538 & 9.6011 &0.0004 &    0.1233 & 0.0004&  $-$7.236 & 0.090&  11.891 & 0.077& 2.170 & 0.068\\
 3 & 5290767150189898496 & 273.949001   & $-$15.745614 & 9.0624 &0.0007 &    0.0739 & 0.0008&  $-$3.940 & 0.048&  10.628 & 0.041& 2.372 & 0.036\\
 4 & 5290673829139963520 & 273.925816   & $-$15.896285 & 9.4731 &0.0008 &    0.2980 & 0.0005&  $-$4.526 & 0.024&   9.926 & 0.025& 2.506 & 0.020\\
 5 & 5290832364972775808 & 273.667326   & $-$15.818870 & 8.3476 &0.0005 & $-$0.0008 & 0.0004&  $-$3.981 & 0.032&  10.661 & 0.028& 2.492 & 0.025\\
 6 & 5290671733195996416 & 273.957022   & $-$15.969642 & 6.8391 &0.0006 &    0.0483 & 0.0005&  $-$4.460 & 0.033&  11.200 & 0.034& 2.463 & 0.027\\
 7 & 5290721108139876864 & 273.869213   & $-$15.886130 & 8.8994 &0.0011 &    0.1359 & 0.0016&  $-$4.160 & 0.025&  13.210 & 0.021& 2.430 & 0.019\\
 8 & 5290722929205920640 & 273.795792   & $-$16.026672 & 8.3232 &0.0005 &    0.0545 & 0.0004&  $-$4.719 & 0.025&  11.331 & 0.023& 2.495 & 0.019\\
 9 & 5290722929205920896 & 273.797944   & $-$16.025092 & 8.7743 &0.0020 &    0.1195 & 0.0021&  $-$4.408 & 0.070&  11.550 & 0.061& 2.648 & 0.049\\
10 & 5290657336465950976 & 274.208409   & $-$16.281212 & 9.3395 &0.0017 &    0.1185 & 0.0020&  $-$4.594 & 0.020&  11.256 & 0.021& 2.464 & 0.017\\
11 & 5290767631226220032 & 273.926845   & $-$15.804975 & 5.8160 &0.0022 & $-$0.1008 & 0.0028&  $-$4.661 & 0.077&  11.226 & 0.063& 2.471 & 0.053\\
\hline
\end{tabular}
\end{table*} 

\begin{figure}
   \centering
\includegraphics[width=0.95\hsize,clip=]{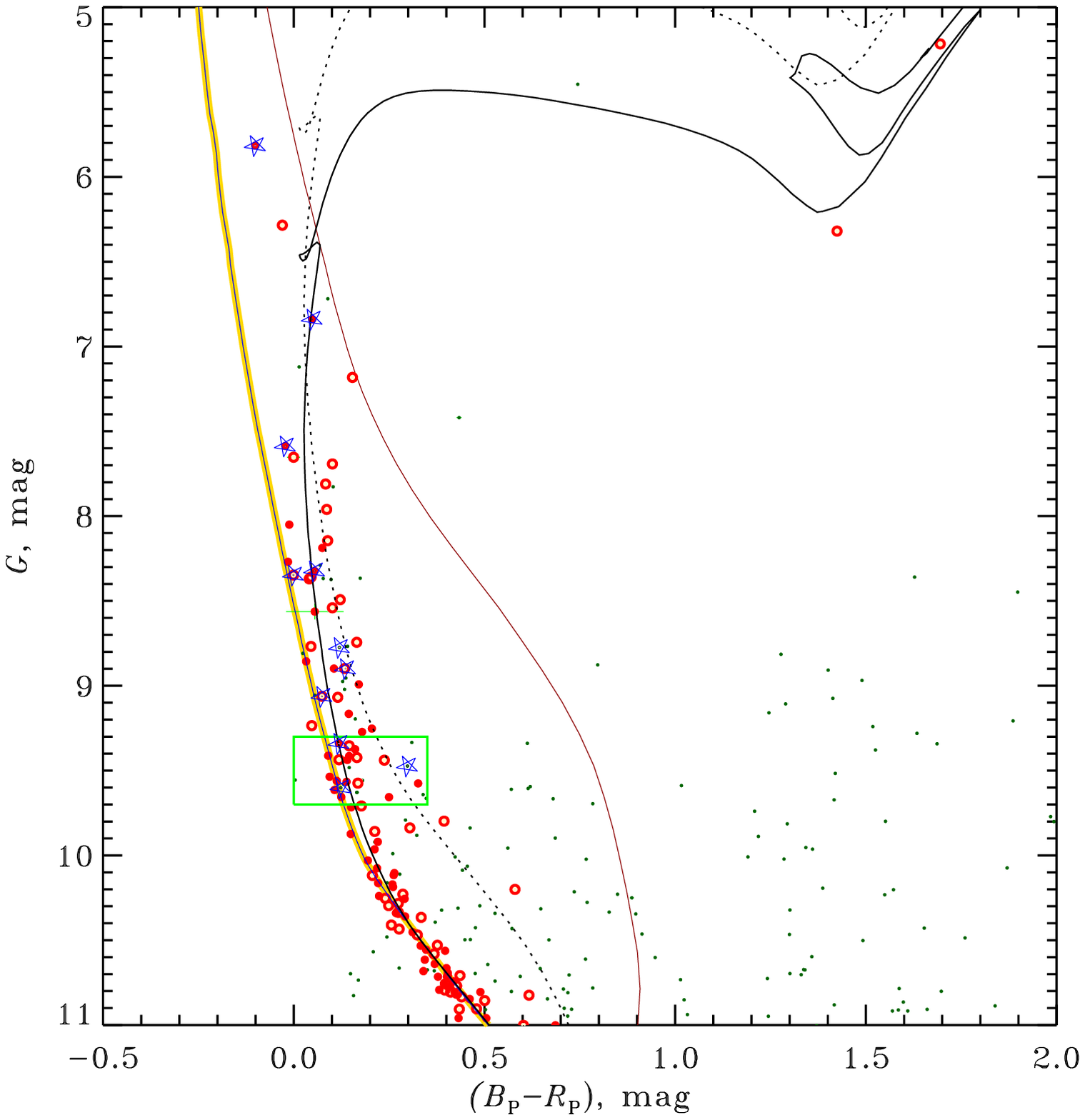}
\includegraphics[width=0.95\hsize,clip=]{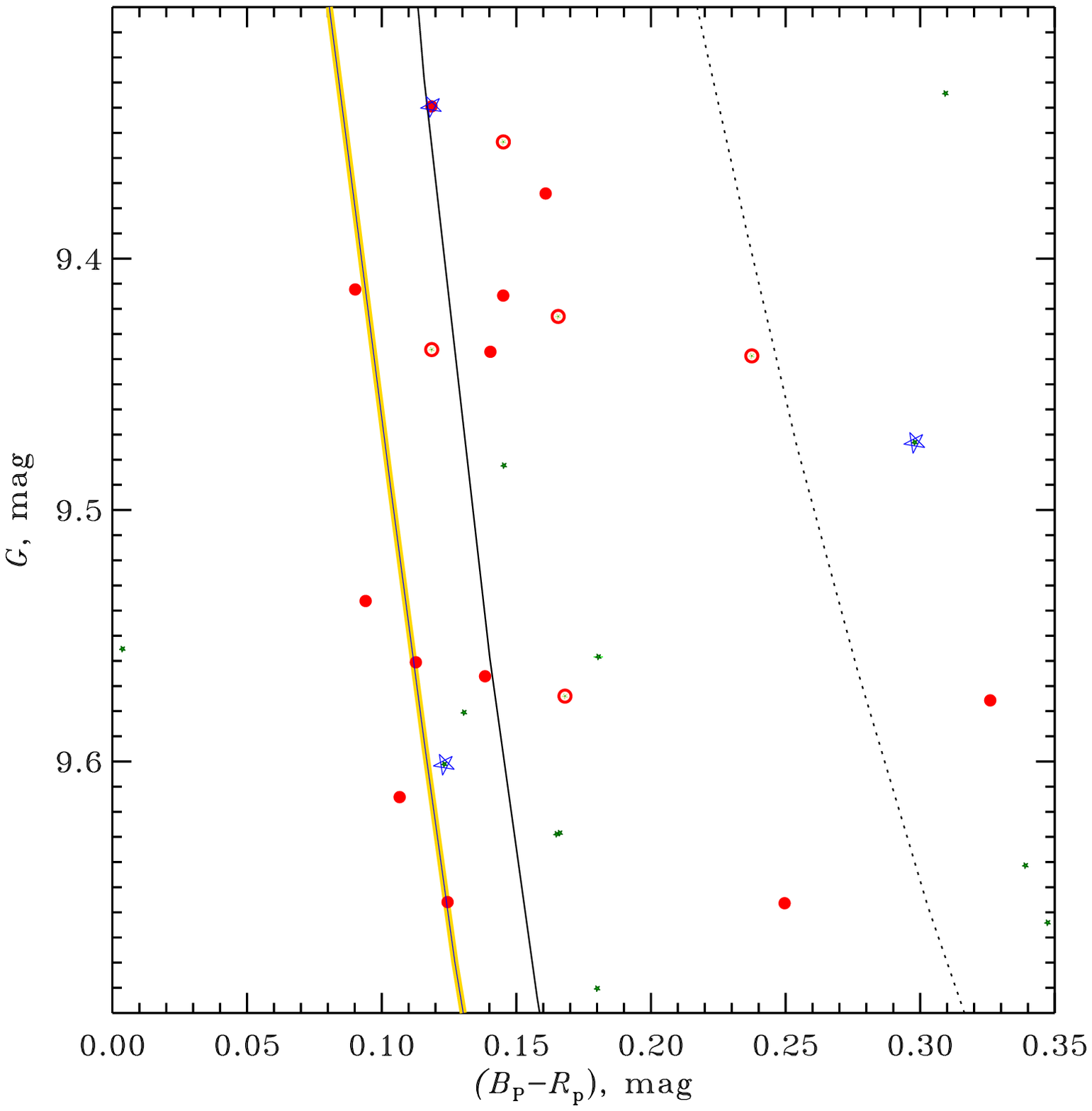}
\caption{
Enlarged version of the CMD shown in Fig.~\ref{fig:cmdmmb} in the Turn Off-region (top panel). Most 
designations are the same as in Fig.~\ref{fig:cmdmmb}. Additionally, we show the stars from 
Table~\ref{tab:cpstars} with blue five-point star symbols. The thin dark red line in the 
top panel shows the TAMS. The green insert is enlarged in the bottom panel.
}
\label{fig:cmdto}
\end{figure}

In Table~\ref{tab:cpstars} we present our sample stars studied for membership in NGC\,2516 and their position in the 
enlarged version of the CMD is shown in Fig.~\ref{fig:cmdto}. The information gathered in this table can be used to 
identify the stars in Fig.~\ref{fig:cmdmmb} and includes their EDR3 identifiers and other relevant data (photometry, 
proper motions, and parallaxes) with their errors. In Table~\ref{tab:cpstares} we present the membership probabilities, 
stellar masses $\log M/M_{\sun}$, ages as $\log t$, and fractional MS-ages $\tau({\rm MS})$. The parameters were interpolated 
over the set of Padova isochrones specified in Sect.~\ref{sec:prm} and covering respective ranges of magnitudes and 
colours. We used for the interpolation our well developed method applied earlier to cluster stars in the projects 
COCD \citep{2005A&A...438.1163K} and MWSC \citep{2012A&A...543A.156K}.
Following \citet{2003A&A...403..645B}, we 
define the fractional age of each star with mass $M/M_{\sun}$ as a fraction of its main sequence lifetime, measured 
from the ZAMS to the TAMS $\tau({\rm MS}) = (t({\rm MS}) - t({\rm ZAMS}))/(t({\rm TAMS})-t({\rm ZAMS}))$.
The errors of the evolutionary parameters 
were propagated from the uncertainties of the stellar positions in the CMD 
$\varepsilon_{G}$, $\varepsilon_{\varpi}$, and $\varepsilon_{(B_{\rm{P}}-R_{\rm{P}})}$ via Monte Carlo simulations.

The spread of the individual ages with respect to the cluster isochrone can have different origins.
Specifically in our case, where the random CMD-placement errors are negligible,
redward shifts ($\log t>8.10$) may reflect the unresolved nature of the source, 
whereas a blueward displacement may be considered as a sign of fast rotation. In the 
case of the brightest MS stars located at the top of the Main Sequence (HD\,66194 and HD\,65987), their younger 
``apparent'' age is interpreted as evidence of the blue straggler nature of the star. In contrast, 
the fractional age reflects the evolutionary advance of the star towards the TAMS.

\begin{table*}

\caption{Membership probabilities of the peculiar stars and their evolutionary parameters computed from the star 
positions in the CMD.
In the first three columns we show the running number, the star name, and the spectral classification given in SIMBAD, followed by  the
membership probabilities $P_{\varpi}$ and $P_{\rm kin}$. In the last six columns, we present stellar mass $\log M$,
age $\log t$, and the fractional MS age $\tau({\rm MS})$, with their respective errors.}
\label{tab:cpstares}
\setlength{\tabcolsep}{4pt}
\centering
\begin{tabular}{rrrrrrrrrrr} 
\hline
\multicolumn{1}{c}{\#}                       &
\multicolumn{1}{c}{Name}                     &
\multicolumn{1}{c}{Spectral}                 &
\multicolumn{1}{c}{$P_{\varpi}$}             &
\multicolumn{1}{c}{$P_{\rm kin}$}            &
\multicolumn{1}{c}{$\log M$}                 &
\multicolumn{1}{c}{$\varepsilon_{\log M}$}   &
\multicolumn{1}{c}{$\log t$}                 &
\multicolumn{1}{c}{$\varepsilon_{\log t}$}   &
\multicolumn{1}{c}{$\tau({\rm MS})$}               &
\multicolumn{1}{c}{$\varepsilon_{\tau({\rm MS})}$} \\
\multicolumn{1}{c}{ }                        &
\multicolumn{1}{c}{ }                        &
\multicolumn{1}{c}{Type}                     &
\multicolumn{2}{c}{ }                        &
\multicolumn{2}{c}{($M_{\sun}$) }            &
\multicolumn{2}{c}{(yrs) }                   &
\multicolumn{2}{c}{ }                        \\
\hline
 1 & HD\,65987        & B9.5IVpSi      &  0.910 &  0.807 & 0.638 &0.002&7.960 &0.007&0.606 &0.010\\ 
 2 & HD\,66318        & A0pEuCrSr      &  0.009 &  0.000 & 0.377 &0.003&8.072 &0.125&0.144 &0.049\\ 
 3 & HD\,66295        & B8/9pSi        &  0.674 &  0.287 & 0.445 &0.002&8.132 &0.031&0.280 &0.021\\ 
 4 & CPD$-$60\,981    & A2Vp:Sr:Cr:Eu: &  0.855 &  0.077 & 0.322 &0.001&8.840 &0.001&0.668 &0.001\\ 
 5 & HD\,65949        & B8/9HgMn       &  0.921 &  0.330 & 0.566 &0.002&7.816 &0.020&0.279 &0.014\\ 
 6 & HD\,65950        & B8IIIHgMn      &  0.995 &  0.950 & 0.627 &0.002&8.163 &0.004&0.944 &0.008\\ 
 7 & CPD$-$60\,978    & A0pEuCrSr      &  0.966 &  0.001 & 0.410 &0.001&8.559 &0.005&0.622 &0.007\\ 
 8 & CPD$-$60\,944\,A & B8pSi          &  0.908 &  0.962 & 0.507 &0.001&8.283 &0.003&0.598 &0.004\\ 
 9 & CPD$-$60\,944\,B & B8III          &  0.111 &  0.750 & 0.428 &0.003&8.516 &0.008&0.629 &0.012\\ 
10 & HD\,65712        & ApSi           &  0.994 &  0.989 & 0.396 &0.002&8.320 &0.018&0.320 &0.014\\ 
11 & HD\,66194        & B3Vn           &  0.984 &  0.998 & 0.851 &0.005&7.614 &0.010&0.886 &0.020\\ 
\hline
\end{tabular}
\end{table*}

 The most striking result of our study of cluster membership is that from the consideration of the 
parallax-proper motion membership, the strongly magnetic A0p star 
HD\,66318 with a mean longitudinal magnetic field $\langle B_z \rangle$ of about 4.5\,kG and  a mean 
magnetic field modulus $\langle B \rangle$ of about 14.5\,kG \citep{2003A&A...403..645B} does not belong 
to the NGC\,2516 cluster at a high level of confidence. 
Using the isochrone for a cluster with an age of 1.6$\times10^{8}$\,yr and the Geneva stellar evolution 
tracks for $Z=0.02$ from \citet{1992A&AS...96..269S}, \citet{2003A&A...403..645B} concluded that HD\,66318 
is at a very young age and has completed only about $16\pm5$\% of its main sequence life. 
The young age of this star was frequently mentioned in the literature as an argument in favour of the 
fossil origin of magnetic fields in chemically peculiar Ap and Bp stars. 
However, it is evident that this star does not belong to NGC\,2516. Thus, its real evolutionary state is undefined. 

Four more chemically peculiar Ap and Bp stars in NGC\,2516 were reported to possess detectable mean longitudinal magnetic 
fields. To investigate the link between the presence of a magnetic field and the evolutionary state in 
chemically peculiar Ap and Bp stars, \citet{2006A&A...450..777B} carried out magnetic field measurements in a 
number of A and B-type stars mentioned in the literature as members of NGC\,2516. The authors reported 
the presence of a magnetic field in the B9p stars HD\,65987, CPD$-$60\,944B, and HD\,66295, and in the 
A0p star HD\,65712. Interestingly, out of these five known magnetic stars, three are reported to be X-ray 
sources \citep{1997MNRAS.287..350J}. The detected X-ray emission could indicate the presence of 
close unresolved lower mass magnetically active companions (e.g.\ \citealt{2001A&A...372..152H}).
According to the work of \citet{2019AJ....157...78J}, who used the Gaia DR2 catalogue to identify 
bright comoving systems in a five-dimensional space (sky position, parallax, and proper motion), all four 
stars  belong either to a comoving binary or multiple stellar candidate systems. In contrast, only 
HD\,65987 has been mentioned to have a companion by \citet{2019A&A...623A..72K} due to the presence of a proper 
motion anomaly detected using the Hipparcos and Gaia DR2 catalogues. \citet{2000AJ....119.2296G} studied the 
cluster membership of the stars HD\,65987, HD\,66295, and CPD$-$60\,944B using astrometric and radial 
velocity data and concluded that HD\,65987 and HD\,66295 are cluster members. 

In our study, out of the five bona-fide magnetic chemically peculiar stars,
only two stars,  HD\,65987 and HD\,65712, have a high membership probability.
Further, the chemically peculiar star CPD$-$60\,944A, the HgMn star HD\,65959, and the blue
straggler candidate HD\,66194 are confirmed as cluster members.
CPD$-$60\,944B can be considered as a proper motion cluster member. Additional information on the targets in our sample
is presented in the Appendix. 

In contrast to the work of \citet{2007A&A...470..685L}, who considered
the magnetic chemically peculiar stars HD\,66295, CPD$-$60\,978, and HD\,66318 as cluster members, our study
shows that all three stars do not fulfil the membership criteria. 
While \citet{2007A&A...470..685L} report that the magnetic chemically peculiar stars HD\,66318 and HD\,65712
have fractional ages below 0.30 -- 0.15 and 0.20, respectively -- and therefore definitely contradict
the work of
\citet{2000ApJ...539..352H,2005ASPC..343..374H}, our results do not confirm cluster membership for
HD\,66318 and indicate a fractional age of 0.320 for HD\,65712. 
                            
The position of the studied chemically peculiar stars in the CMD of NGC\,2516 is presented in the upper panel of Fig.\ref{fig:cmdto}.
The obtained main-sequence fractional ages for the bona fide ($1\sigma$) cluster members range from 0.320 for 
the fainter chemically peculiar magnetic star HD\,65712 to 0.944 as determined for the second brightest
HgMn star in the sample, HD\,65950. 
The brightest star, HD\,66194, with a fractional age of 0.886, demonstrates an individual age significantly younger than 
the common cluster age of $\log t=8.10$. 
HD\,65987 with a fractional age of 0.606 also appears to be younger than the cluster itself. The 
apparent younger ages of HD\,66194 and HD\,65987 suggest that they are associated with stellar merging, where a merge or mass transfer 
took place.

Among the chemically peculiar stars with confirmed cluster membership, the strongest mean longitudinal magnetic field 
was measured for HD\,65712 ($\langle B_z \rangle=1.1\pm0.05$\,kG) showing the lowest fractional age of 0.320. 
The second strongest mean longitudinal magnetic field ($\langle B_z \rangle=0.6\pm0.1$\,kG) was measured for 
HD\,65987 with $\tau({\rm MS})=0.606$, and the third strongest field ($\langle B_z \rangle=0.35\pm0.06$\,kG) was 
determined for CPD$-$60\,944B with a fractional age of 0.629. The fact that the strongest magnetic field was 
detected in a star with the lowest fractional age is, however, difficult to interpret in terms of stellar 
evolution effects, as our sample is much too small to allow the deduction of any statistical evidence.

\section{Discussion}
\label{sect:conc}

In this study, we have aimed at using the most accurate and complete Gaia EDR3 data on stellar astrometry and  
photometry in the area of the nearby rich open cluster NGC~2516 to establish the membership probability of 
known peculiar stars and to deduce the evolutionary state of these stars. Since the stars reside in the 
central area of the cluster, we have confined ourselves with the consideration of the inner part of the cluster 
with a radius of 1\,deg and selected 37508 stars brighter than $G=19$\,mag. To determine their membership 
probabilities, we used our tried-and-true approach already applied for the COCD and MWSC surveys. The  
only difference between the presented and previous studies is that we do not use the photometric data for membership evaluation.

As a byproduct, we have determined the average astrometric parameters 
of the cluster (the proper motion and parallax), allowing the accurate placement of the cluster stars in the CMD.  
In order to use effectively the cluster CMD, we applied the Padova isochrones
for the determination of the evolutionary parameters.
The high precision of the EDR3 data allowed us to employ the full range of the available magnitudes 
from the very top of the cluster CMD down to the faintest stars. The modest variations of the isochrone metallicity 
and the used transformation scale around the expected values enabled us to find the best combination of the model 
parameters providing the best fit of the isochrone and the observations in the $G,B_{\mathrm{P}}-R_{\mathrm{P}}$  
plane. We find that the isochrone perfectly fits the observed cluster locus, both in the brightest 
red giant domain at $G\approx5$\,mag, and at the faintest magnitudes $G\approx 18$ (with a marginally visible 
pre-MS branch). In spite of the perfect overall agreement, we have to note some disaccord of 
$\Delta(B_{\mathrm{P}}-R_{\mathrm{P}})\lesssim 0.1$ between the  model and the observational colours for 
intermediate magnitudes ($G\simeq12-16$). Note that both the upper CMD of the selected members and its 
lower part agree well with an isochrone of $\log t=8.10$.
The derived cluster parameters allowed us to safely discriminate in parallax-proper motion space between 
foreground and background field stars and probable cluster members. In total, we found in the queried sample 719 
probable ($P_{c}>0.61$) and 764 possible ($P_{c}=0.14-0.61$) members of the cluster. 

The parallax-proper motion 
membership pipeline indicates that only 5 of the 11 considered CP stars (HD\,65987, HD\,65950, 
CPD$-$60\,944\,A, HD\,65712, and HD\,66194) should be classified as highly probable members, and three more 
(HD\,66295 and HD\,65949, and CPD$-$60\,944B) as possible members. The remaining CP stars are definite field stars, being 
incompatible with the cluster kinematics (HD\,66318, CPD$-$60\,981, and CPD$-$60\,978).

In general, the Ap stars of our sample are bright enough ($7\lesssim G\lesssim 10$\,mag) and appear to be excellent 
targets from the formal point Gaia EDR3 observations \citep{2021A&A...649A...1G,2021A&A...649A...3R,2021A&A...649A...5F} 
to provide a robust estimation of their membership probability. One should, however, keep in mind that their physical 
nature and their environment may affect the derived parameters and change the reached conclusions. Indeed, these stars 
are immersed as a rule in a dense stellar field and  several targets in our sample are binary or multiple systems. As an 
example of such a complication, we can mention CPD$-$60\,944, which according to the CCDM catalogue \citep{2002yCat.1274....0D} 
is a quadruple system with bright ($V\sim9-11$\,mag) components located at a distance of 9 to 46\,arcsec from the primary.  
Only for the A and B components the Gaia EDR3 is able to provide data. However, the Gaia data obtained for the A and B 
components of about the same apparent magnitude $G$ show rather different proper motions and parallaxes (only marginally 
compatible to each other) and are accompanied with twice as large errors. The high parallax membership probability for the 
primary, and the low probability for the secondary leads to only a small chance for the secondary to belong to the cluster. 
As a result, the combined membership probability $P_{c}$ of the secondary falls below the possible cluster member threshold 
adopted in this study. This indicates that a statistical approach does not work for this star, due to either an error in the 
parallax determination or to a wrong classification as a physical companion of the primary. Currently, we are inclined to 
regard this star to be a proper motion cluster member with complicated parallax data. Therefore, for this, as well as some 
other difficult cases (e.g.\ for HD\,65949, which is a triple system) one should treat the derived probabilities with 
caution and consider additional arguments such as photometry, spatial position, radial velocities (expected soon with the 
Gaia DR3 release), to establish their membership more reliably.

The possible blue straggler nature of HD\,65987 was already mentioned by \citet{1995A&AS..109..375A} and is discussed 
here again based on the high precise Gaia EDR3 data \citep{2021A&A...649A...1G}. \citet{1995A&AS..109..375A} found blue 
stragglers in all clusters of all ages and the percentage of clusters with blue stragglers grows with age and richness 
of the cluster. In NGC\,2516, HD\,66194 and HD\,65987 are clearly located to the left of the cluster main sequence isochrone 
and can be considered as blue stragglers. As already mentioned in Sect.~\ref{sec:indiv}, HD\,65987 is reported to have 
a companion by \citet{2019A&A...623A..72K} due to the presence of a proper motion anomaly detected using the Hipparcos 
and Gaia DR2 catalogues. Also  \citet{2000AJ....119.2296G} reported on variations in the line profiles, suggesting the 
probable presence of a companion. To our knowledge, HD\,65987 is the first chemically peculiar blue straggler with a 
detected magnetic field of the order of a few hundred Gauss. No significant magnetic field was detected in the previous 
rare attempts to measure a  magnetic field in blue stragglers belonging to other clusters and associations 
(e.g.\ \citealt{2008A&A...488..287H}).  According to \citet{2008A&A...488..287H}, the star $\theta$~Car, a blue straggler 
in the open cluster IC\,2602, was rejuvenated due to a previous mass-transfer episode. The results of the search for a 
magnetic field using FORS 1 at the VLT consisting of 26 measurements over a time span of 1.2\,h have been rather 
inconclusive with only a few measurements having a significance level of 3$\sigma$.

As mentioned in Sect.~\ref{sec:intro}, the mechanism of the generation and the maintenance of magnetic fields in
chemically peculiar stars is not well understood yet, although the currently most popular scenario is that 
magnetic fields may be generated by strong binary interaction. Therefore, with respect to the various scenarios
for the magnetic field origin in upper main sequence stars, the confirmation of the blue straggler status of
the chemically peculiar magnetic star HD\,65987 is of great importance. Obviously, future magnetic studies should
involve a search for magnetic fields in astrometrically confirmed blue stragglers to support the suggested scenario.

\section*{Acknowledgements}

This work has made use of data from the European Space Agency (ESA) mission {\it Gaia} (\url{https://www.cosmos.esa.int/gaia}), 
processed by the {\it Gaia} Data Processing and Analysis Consortium (DPAC, \url{https://www.cosmos.esa.int/web/gaia/dpac/consortium}). 
Funding for the DPAC has been provided by national institutions, in particular the institutions participating in the {\it Gaia} 
Multilateral Agreement. We also made use of the SIMBAD database, operated at CDS, Strasbourg, France, and of the WEBDA 
open cluster database. The use of \textit{topcat}, an interactive graphical viewer and editor for tabular 
data \citep{2005ASPC..347...29T}, is acknowledged.

\section*{Data Availability}

The data used in this study was queried with the \textit{topcat} facility from the Gaia EDR3 catalogue 
(tables \texttt{gaiaedr3.gaia\_source} and \texttt{gaiaedr3.gaia\_source\_corrections}) at Astronomisches 
Rechen-Institut, Heidelberg, Germany (site \texttt{https://gaia.ari.uni-heidelberg.de/tap}) using the 
following ADQL query\\
\vskip1mm
\noindent
\texttt{SELECT  \\
   TOP 100000   \\
   *  \\
   FROM gaiaedr3.gaia\_source AS edr \\
   JOIN \\
      gaiaedr3.gaia\_source\_corrections USING (source\_id ) \\
   WHERE \\
     1=CONTAINS(POINT('ICRS', edr.ra, edr.dec), \\
                CIRCLE('ICRS', 119.490, -60.750, 1.000 ))
}\\                

\appendix
\section{Additional information on individual targets}
\vskip1mm

{\it CPD$-$60\,944B:} This star was reported to be a member of a visual pair with the visual companion  CPD$-$60\,944A,
with an undetected magnetic field \citep{2006A&A...450..777B}. It exhibits in its spectrum strong
lines of \ion{Hg}{ii}, \ion{Mn}{ii}, \ion{P}{ii}, \ion{Ga}{ii}, and \ion{Xe}{ii}
and was identified as a HgMn star \citep{2014MNRAS.443.1523G}. 
Radial velocity measurements of CPD$-$60\,944A carried out by \citet{2000AJ....119.2296G} suggested the 
presence of an additional companion, making CPD$-$60\,944A a triple system. However, according to the CCDM 
catalogue\footnote{https://cdsarc.cds.unistra.fr/viz-bin/cat/I/274} CPD$-$60\,944\,A and CPD$-$60\,944\,B belong
to a quadruple system.
Based solely on its low parallax membership probability, 
the magnetic B9p star CPD$-$60\,944B seems not be a member of the cluster NGC\,2516.
However, CPD$-$60\,944B shows sufficient kinematical probability to be a safe cluster member,
which is in agreement with the study by 
\citet{2000AJ....119.2296G}. Currently, we consider this star to be a proper motion cluster member.
Due to the complex nature of this system it is necessary to use additional criteria
for the membership like photometry, spatial position, and radial velocities. 

{\it CPD$-$60\,978:} The chemical peculiarities in the atmosphere of CPD$-$60\,978, typical for magnetic Ap stars, were mentioned 
for the first time by \citet{1972A&A....21..373D}. This star was also reported as an X-ray emission source. 
No definite detection of a magnetic field was reported by \citet{2006A&A...450..777B}. In contrast to the 
results of the study by \citet{2000AJ....119.2296G}, who identify this star as a cluster member, our work 
shows a very low $P_{\rm kin}$.

{\it CPD$-$60\,981:} Our study shows that CPD$-$60\,981, with a membership probability $P_{\rm kin}$ of only 7.7\% is a possible 
field star according to our classification criteria.
This star is a short-period eclipsing binary with an orbital period $P_{\rm orb}=3.2$\,d \citep{2001A&A...374..204D}
and was classified as an Ap star with 
SrCrEu peculiarity type by \citet{1976ApJ...205..807H}. \citet{1981A&A....96..151M} reported on the measurement 
of a small value of the photometric $\delta$a index, indicating that this system might host a magnetic Ap star. 
However, \citet{2001A&A...374..204D} were not able to detect any strong peculiarity of Ap type and concluded 
that any peculiarity, if present, can only be very mild.  No detection of a magnetic field was achieved by 
\citet{2006A&A...450..777B}. 

{\it HD\,66194:} The blue straggler nature of the fast rotating B3Vn star HD\,66194 was discussed at length by 
\citet{2000AJ....119.2296G}. It is a photometric variable of $\gamma$\,Cas-type and the spectrum of this 
star presents Balmer emission lines. This star was mentioned to have a companion by 
\citet{2019A&A...623A..72K} due to the presence of a proper motion anomaly detected using the Hipparcos 
and Gaia DR2 catalogues. The high rotation rate of this star may indicate the presence of mass transfer 
in a close binary system. No magnetic field was detected in this star by \citet{2006A&A...450..777B} 
using low-resolution spectropolarimetry. This target was reported as the strongest X-ray source in the X-ray cluster survey 
carried out by \citet{1997MNRAS.287..350J}. Our results confirm membership of this star in NGC\,2516.

{\it HD\,65950 and HD\,65949:} The presence of a weak magnetic field was detected in the two HgMn stars HD\,65950 and HD\,65949 by 
\citet{2006AN....327..289H} using low-resolution FORS\,1 spectropolarimetry. A weak mean longitudinal magnetic field
was detected in HD\,65949 also using high-resolution HARPS spectropolarimetry 
\citep{2020MNRAS.495L..97H}. In our study only HD\,65950 is confirmed to be a bona-fide member of NGC\,2516, in accordance 
with the work of \citet{2000AJ....119.2296G}. HD\,65949 was suggested to be a triple system due to a 
small variation in the centre-of-mass velocity, interpreted as due to the presence of a third body 
\citep{2010MNRAS.405.1271C}. Also HD\,65950 is not a single star, as it was reported to have a companion 
due to the presence of a proper motion anomaly, detected using the Hipparcos and Gaia DR2 catalogues 
\citep{2019A&A...623A..72K}. In contrast to the chemically peculiar Ap and late Bp stars, which are rarely 
members of close binaries, but rather frequently members of 
wide systems \citep{2017A&A...601A..14M},  HgMn stars predominantly appear in binary and multiple systems 
\citep{1995ComAp..18..167H,2010A&A...522A..85S,2020MNRAS.496..832C}, in particular in close binaries 
with orbital periods between 3 and 20\,d \citep{1995ComAp..18..167H,2020MNRAS.495L..97H}.

\bibliographystyle{mnras}
\bibliography{ngc2516}

\bsp	
\label{lastpage}
\end{document}